\documentclass[11pt,citesort]{article}
\usepackage[T1]{fontenc}
\usepackage[latin9]{inputenc}
\usepackage{units}
\usepackage{graphicx}
\usepackage{esint}
\usepackage{color}

\newcommand\be{\begin{equation}}
\newcommand\ee{\end{equation}}
\newcommand\bea{\begin{eqnarray}}
\newcommand\eea{\end{eqnarray}}
\newcommand{\bdm}{\begin{displaymath}}
\newcommand{\edm}{\end{displaymath}}
\newcommand\nn{ \nonumber\\}

\begin{document}

\title{String-Gauge Dual Description  of Deep Inelastic Scattering  \\ at Small-$x$}

\author{Richard C. Brower%
\thanks{Physics Department, Boston University, Boston MA 02215%
},
Marko Djuri\'{c}%
\thanks{Physics Department, Brown University, Providence RI 02912%
},
Ina Sar\v{c}evi\'{c}%
\thanks{Physics Department, University of Arizona, Tucson AZ 85721%},
}
\thanks{Department of Astronomy and Steward Observatory, University of Arizona, Tucson, AZ 85721%
}, 
and
 Chung-I Tan%
\thanks{Physics Department, Brown University, Providence RI 02912 %
} }
\maketitle
\begin{abstract}
The AdS/CFT correspondence in principle gives a new approach to deep
inelastic scattering as formulated by Polchinski and
Strassler. Subsequently Brower, Polchinski, Strassler and Tan  (BPST)
computed the strong coupling kernel for the vacuum (or Pomeron)
contribution to total cross sections. By identifying deep inelastic scattering
with virtual photon total cross section, this allows a self consistent description
at small-$x$ where the dominant  contribution
is the vacuum exchange process. Here we formulate this contribution
and compare it with HERA small-$x$  DIS scattering data.
We find that the BPST kernel along with a very simple local approximation to
the proton and current ``wave functions'' gives a remarkably  good fit
not only at large $Q^2$ dominated by conformal symmetry but also extends to small
$Q^2$, supplemented by a hard-wall cut-off of the AdS in the IR. We suggest that
this is a useful phenomenological
parametrization with implications  for other  diffractive processes, 
such  as double diffractive Higgs production.
\end{abstract}

\newpage

\section{Introduction}

Deep inelastic scattering (DIS) provides the primary experimental probe for determining the structure of the proton, which is of vital importance to the interpreations and callibration of the experimental program at the Large Hardron Collider (LHC). However the usefulness of DIS data is limited by the difficulty of a deep theoretical understanding of Quantum Chromodynamics in this regime, notwithstanding the progress made in  the ``small-x'' physics from the perspective of perturbative QCD~\cite{Gribov:1983}, i.e., the so-called BFKL Pomeron~~\cite{Lipatov:1996ts,Lipatov:1976zz,Kuraev:1977fs,BL}.  The problem is that even in this regime the phenomenon is a subtle mixture of perturbative and non-perturbative physics.  Recently the AdS/CFT (or more generally string/gauge duality) suggests a new approach to DIS at strong coupling starting from the paper by Polchinski and Strassler \cite{Polchinski:2002jw}. This article begins to explore the possibility of developing a new phenomenological framework based on the application of strong coupling Pomeron kernel computed by Brower, Polchinski, Strassler and Tan (BPST)~\cite{Brower:2006ea,Brower:2007qh,Brower:2007xg}.   (For other related works, see \cite{Cornalba:2006xk,Cornalba:2006xm,Cornalba:2007zb,Cornalba:2007fs}.) By testing and calibrating the phenomenology of the BPST kernel against the DIS scattering data of HERA at small-x, we hope to provide a foundation for extending this approach to higher energies and to other diffrative processes, such as an estimation of the rate of double-diffractive  Higgs production at the LHC.

Before introducing the AdS/CFT strong coupling approach in
Sec.~\ref{sec:BPSTformalism}, it is useful to make some general
observations on how we treat Pomeron physics.  In AdS/CFT,
non-perturbative physics is organized following the original
observation of 't Hooft. Namely, QCD can be expanded (formally) term
by term as a power series in  $1/N_c$ at fixed 't Hooft coupling
$\lambda = g_{YM}^2{N_c}$. As a consequence, various non-perturbative
effects are classified in terms of a topological (or string theoretic)
expansion. This has many well known qualitative successes. For
example, the leading term for mesons is  the valence approximation and
for scattering Zweig-rule violating processes are suppressed. The
nucleon is introduced as an external probe after we set $N_c$ to its physical value, $N_c =3$. At high energies the vacuum exchange in leading order of $1/N_c$-expansion  is the cylinder diagram, which unambiguously {\bf defines} the ``elementary'' Pomeron as a non-perturbative color singlet gluonic object. This is in fact completely consistent with the weak coupling BFKL Pomeron, which is the leading large $N_c$ contribution to first order in the 't Hooft coupling $g^2N_c$ and all orders in $g^2 N_c \log\;s$. The BFKL equation can be viewed as  the ladder approximation in the color singlet two-gluon channel.

\begin{figure}[htpb]
\begin{center}
\includegraphics[width=.6\textwidth]{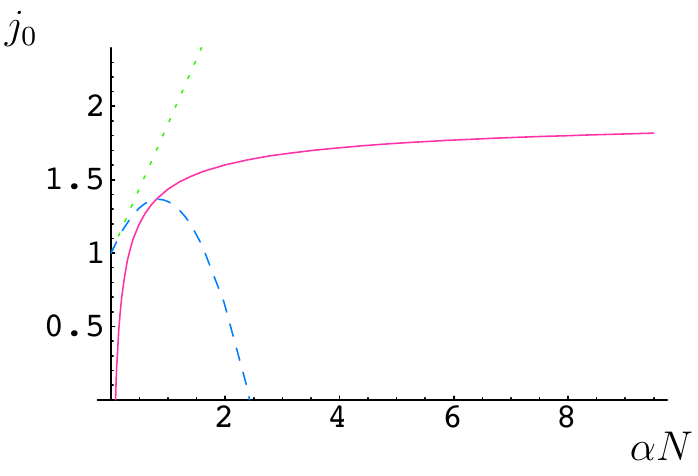}
\caption{The Pomeron intercept in QCD viewed as a function of the 't Hooft coupling. ($\lambda=g_{YM}^2N_c$, with $\alpha=g^2_{YM}/4\pi$.) This is reproduced from Fig. 12 of Ref. \cite{Brower:2006ea}.}
\label{fig:Interceptb}
\end{center}
\end{figure}

In the strong coupling limit, BPST Pomeron is now computed
non-perturbatively in the $1/N_c$ expansion in leading order in
$\lambda = 1/g_{YM}^2N_c$ at strong coupling, with an intercept 
\be
j_0= 2-2/\sqrt \lambda + O(1/\lambda)\; . \label{eq:BPST}
\ee
 Thus the two gluon weak coupling BFKL
Pomeron is now viewed as Reggeized Graviton in a confining AdS
background. In spite of the change of view point, the qualiative features of weak and strong coupling kernels are strictly similar, due no doubt to lack of conformal breaking in their resepective approximations.  It is instructive to compare the value of the leading power $j_0$ for forward scattering for both strong and weak couping as given in Fig.~\ref{fig:Interceptb}. Observe that a power of $j_0 \simeq 1.3$ lies  securely in the cross over region between strong and weak coupling.

Turning to DIS scattering, (e.g. the total cross section for a virtual Compton  scattering at fixed $Q^2$), we now have a further tool to probe the Pomeron kernel. Varying the virtual photon's momentum, in BFKL language, $1/Q$ probes the size of the two gluon dipole, whereas from AdS/CFT dictionary, it probes the 5th radial coordinate, $z \simeq 1/Q$, in the AdS space.  Here we can ask how the BPST Pomeron can explain  the observed $Q^2$-dependence for the effective power behavior for virtual Compton scattering
at fixed $Q^2$ as seen in the HERA data~\cite{Breitweg:1998dz}. ( See Fig. \ref{fig:Intercepta}. Note that the effective power increases with $Q^2$.)

It is important to point out  that the strong coupling approach provides
a natural way to  include the non-conformal contributions due to confinement by a deformation to the $AdS_5$ geometry. Thus the hard-wall
AdS  Pomeron  provides  a synthesis of the so called ``soft Pomeron'', i.e.,  a Regge
pole which interpolates with a tensor glueball at
$j = 2$ and the ``hard Pomeron'', characterized by the BFKL weak coupling
behavior.

The main result of this paper is to show how well the exchange of a single strong coupling Pomeron describes the HERA data for the small-x region of DIS. (Other recent studies include \cite{BallonBayona:2007qr,BallonBayona:2007rs,Cornalba:2008sp,Cornalba:2009ax,Cornalba:2010vk,Hatta:2007cs,Hatta:2007he,Hatta:2008zz,Albacete:2008dz,Albacete:2008ze,Kovchegov:2009yj,Kovchegov:2010uk,Taliotis:2009ne}.) Indeed when the hard-wall model is introduced to implement confinement the fit is remarkably good even down to $Q^2 =0.1 \; GeV^2$.  In the $1/N_c$ expansion, there are of course non-linear effects which enter through eikonalization. When this effect becomes important for DIS, it can be interpreted as the onset of ``saturation''.  We determine that for the range of energies given by HERA data these effects are still negligible for $Q^2\geq O(1) \; GeV^2$, but that they will come into play at LHC energies.

The organization of the paper is as follows.  In Sec.~\ref{sec:background}, we begin with a short discussion on the kinematics for the HERA small-x regime and the standard weak-coupling partonic approach to DIS. We next briefly review the basic AdS/CFT formalism.  In Sec.~\ref{sec:BPSTformalism} we treat DIS in the small-x limit to first order in the BPST Pomeron. We begin first with the conformal limit, and next discuss the modification due to confinement, using the hard-wall model as an illustration.  We clarify how at strong coupling the small-x Regge limit and the large-$Q^2$ limit are unified. In  Sec.~\ref{sec:linearfit} we test our strong coupling results by fitting to the recently combined H1-ZEUS small-x data from HERA~\cite{:2009wt}. We focus on a single-Pomeron contribution based on a local approximation for both the current and the proton ``wave functions''. We find that the confinement-improved treatment (hard-wall model) allows a surprisingly good fit to all HERA small-x data, with $Q^2$ ranging from $0.1\, GeV^{2}$ to $400\, GeV^{2}$.  The single-Pomeron hard-wall fit also indicates possible onset of ``saturation'' for $Q^2 \leq O(1) \; GeV^2$. In Sec.~\ref{sec:Saturation} we carry out a nonlinear eikonal analysis.  It is now important to fully explore the dependence of the eikonal, $\chi( s, \vec b, z,z')$, on the 3-dimensional transverse space, i.e., $\vec b$ and $z$.  Our analysis confirms that saturation effect is small for $Q^2\geq O(1) \; GeV^2$ at HERA energy range. However, for $Q\leq O(1) \; GeV^2$, confinement-improved eikonal treatment allows a better fit to the HERA small-x data. Finally, we summarize in Sec.  \ref{sec:discussion} our findings and discuss their possible significance.

{\it Note Added:} After initial posting of this work, we became aware of Ref. {\cite{Levin:2010gc} which has also confirmed our prior claim, demonstrated more explicitly here, that,  for the HERA range, saturation effect is minimal.

\section{Background}\label{sec:background}

\subsection{Parton Phenomenology}\label{sec:parton}

In a partonic approach, hadron structure functions $F_i(x,Q^2)$ in DIS follow from quark and gluon distributions, $q_i(x,Q^2)$, $\bar q_j(x,Q^2)$, and $g_k(x,Q^2)$  respectively. At small-$x$, QCD evolution dictates that most of the wee partons  evolve from gluons and the valence contributions become subdominant. At LHC, proton gluon distribution is expected to play an increasingly important  role at small-$x$ and/or large $Q^2$.  In a strong coupling approach, although it is no longer meaningful to speak of quarks and gluons as the effective partonic degrees of freedom, the dominance of gluon dynamics at small-$x$ justifies  treating a simpler situation where quarks are first ignored, i.e., considering  the limit $N_c>>N_f$. In such a limit, Gauge/String duality can be applied directly~\cite{Polchinski:2002jw,Brower:2006ea}.

\begin{figure}[bthp]
\begin{center}

\includegraphics[width=.8\textwidth]{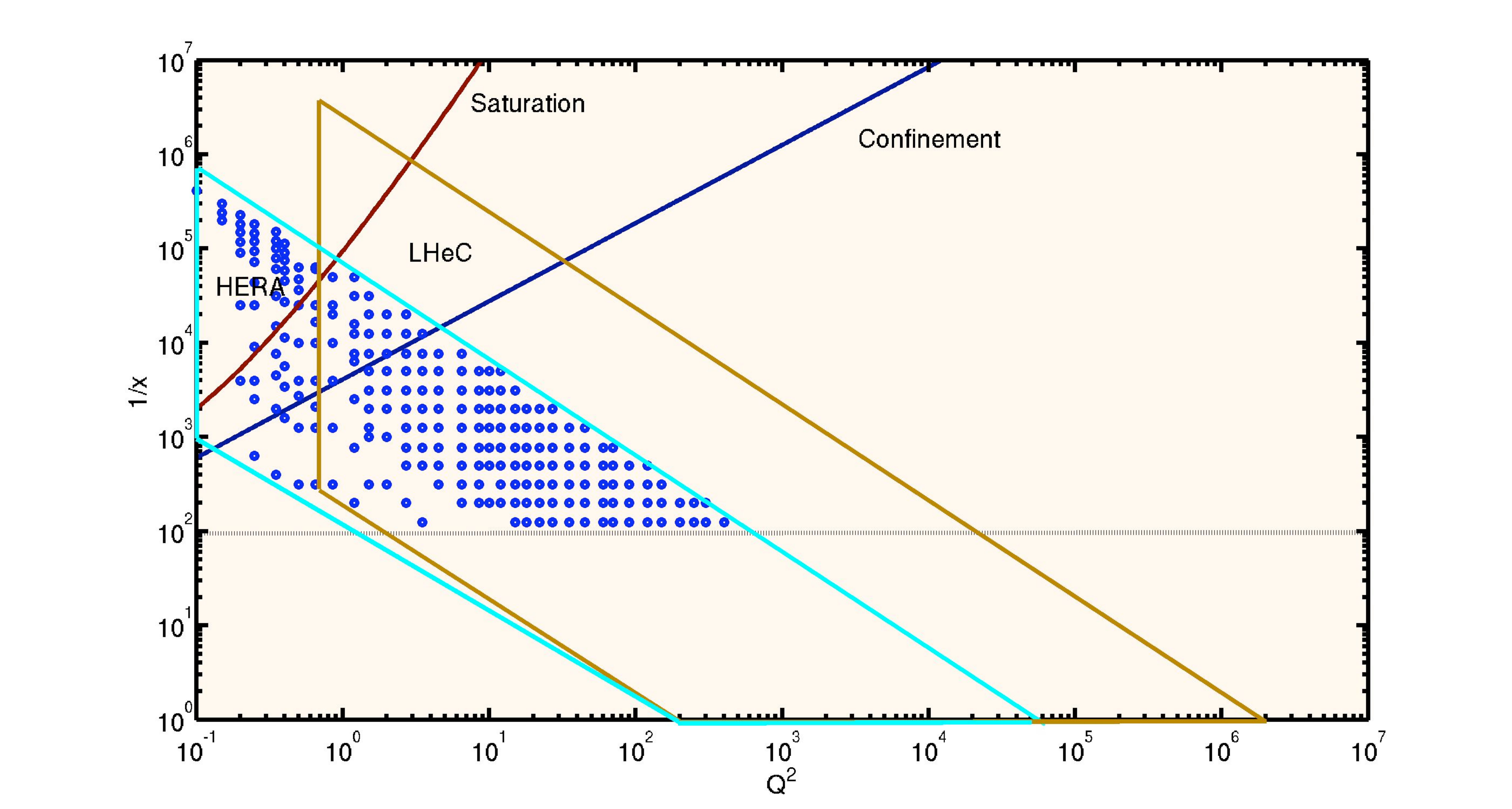}

\caption{  HERA and LHeC regions are labeled by trapezoids   respectively.  We have also shown the entire combined H1-ZEUS small-$x$ data points as dots. Lines for confinement  and saturation are discussed in Sec. \ref{sec:BPSTformalism} and Sec. \ref{sec:Saturation} respectively.}
\label{fig:HERA_LHC}
\end{center}
\end{figure}

In  a naive parton picture, Bjorken scaling would lead to $Q^2$-independence for the DIS structure functions.  In a perturbative treatment, due to QCD evolution, structure functions such as $F_2(x, Q^2)$ become $Q^2$-dependent,  which can be derived by using the linear DGLAP evolution equation. This has indeed been carried out for the HERA data, starting with phenomenological inputs for structure functions at low $Q^2$, e.g., $Q^2 \simeq 3-4 \; GeV^2$. Of course, an inclusive  DIS cross section  can also be treated   directly as   the total cross section of an off-shell photon, with  mass squared $-Q^2$, scattered off a proton. The Bjorken $x$ variable  is then related to the center of mass energy squared $s$ of the $\gamma^*p$ system by $x\approx\nicefrac{Q^{2}}{s}$, at high $s$. The limit $x \rightarrow 0$, with $Q^2$ fixed, can then be considered as the Regge limit for $\gamma^*p$ two-body scattering.   The off-shell $\gamma^*p$ total cross section in the limit $x\rightarrow 0$   can then be characterized by  the exchange of Pomerons, as done for other hadronic total cross sections. This has been studied at
weak coupling using the BFKL equation, which corresponds to  summing over Reggeized gluon ladder graphs \cite{Lipatov:1996ts,Lipatov:1976zz,Kuraev:1977fs,BL}.
It is therefore tempting to identify the rise of $F_2(x,Q^2)$ with $1/x$, for fixed $Q^2$, observed at HERA,  as the onset of the Pomeron dynamics. In Fig. \ref{fig:HERA_LHC}, we indicate in a $\log (Q^2)-\log (1/x)$  plot the kinematic  region covered by HERA and also the region of particular interest at LHC, e.g., that for the proposed LHeC. We also identify, by dots, the small-$x$ data points, with $x < 0.01$,  for  the combined H1-ZEUS data set, which has been released recently~\cite{:2009wt}.  There are 249 data points, with $Q^2$ ranging from $0.1$ to $400\; GeV^2$.

DGLAP provides a description for the  evolution of the data in $Q^2$, given $F_i(x,{Q}^{*2})$ at some initial point ${Q}^{*2}$. The study of small-$x$, on the other hand, focuses on evolution in $1/x$, with $Q^2$ fixed.  Indeed, for $Q^2$ ranging from $0.15\, GeV^{2}$ to $250\, GeV^{2}$,  by fitting the data to
\be
F_2(x, Q^2) \sim (1/x)^{\epsilon_{eff}},
\ee
with $Q^2$ fixed,   ``effective Pomeron" intercept, $1+\epsilon_{eff}$, can be extracted~\cite{Breitweg:1998dz}, with $\epsilon_{eff}$ shown in Fig. \ref{fig:Intercepta}.  There are two  intriguing aspects to this analysis.  First, the effective intercept, except for that at small $Q^2$, is very large, with $1+\epsilon_{eff}\simeq 1.2 \sim 1.4$.
Even more puzzling is the fact that $\epsilon_{eff}$ is $Q^2$-dependent, contrary to a naive Regge expectation. This feature  could seriously challenge the hypothesis of Pomeron dynamics at work in the  HERA energy range. Indeed, this $Q^2$-dependence has been used to support the notion of ``two-Pomeron'' hypothesis~\cite{Donnachie:1998gm}. (For related earlier work, see \cite{Capella:1992yb,levintan,Bondarenko:2003xb}.)

As we have mentioned earlier, using AdS/CFT,  the BPST Pomeron can provide a synthesis of both the ``soft'' and the ``hard''  Pomerons. Using the BPST Pomeron, we have found that the $Q^2$-dependence for $\epsilon_{eff}$ observed at HERA, Fig. \ref{fig:Intercepta}, can be attributed primarily to diffusion for $Q^2$ large and to confinement effects for $Q^2$ small.

\begin{figure}[bthp]
\begin{center}
\includegraphics[width=0.7\textwidth]{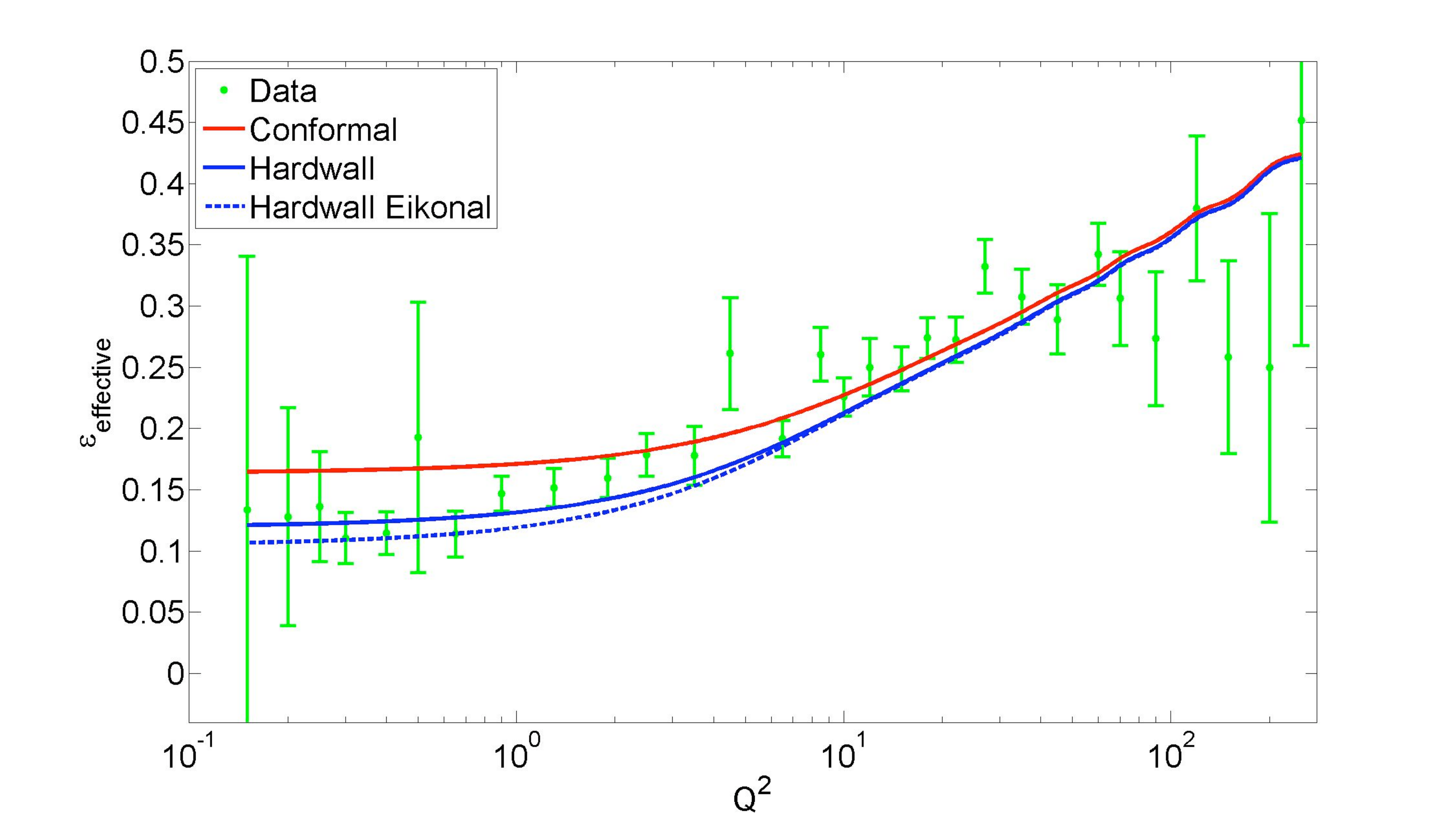}
\caption{$Q^2$-dependence for effective Pomeron intercept,
$\alpha_{P}=1+\epsilon_{eff}$,  from \cite{Breitweg:1998dz}.  Various curves are fits  based on strong/gauge duality described in Secs. \ref{sec:linearfit} and \ref{sec:Saturation}.}
\label{fig:Intercepta}
\end{center}
\end{figure}

\subsection{Reveiw of AdS/CFT Formalism}\label{sec:PSformalism}

In this paper, we will re-examine the DIS structure functions at
small-$x$ by treating the off-shell ${\gamma^*p}$ total cross section
using the  AdS/CFT correspondence. That is, with String/Gauge Duality, we can provide a direct
strong coupling treatment for the DIS structure functions, consistent with DGLAP expectation. We focus on the conformal limit as well as
effects of confinement and evidence for the onset of
saturation.  We support the usefulness of this strong coupling
formalism by testing against the recently published combined H1-ZEUS
data set.  Extrapolation to the LHC kinematic region will also be
provided.

In our approach, two-body amplitudes are controlled by exchanging BPST Pomeron,  with an intercept
\be
j_0\simeq 2 - \rho
\ee
where we have re-written Eq. (\ref{eq:BPST}) in terms of a more convenient parameter, $\rho\equiv 2/\sqrt \lambda$,  $\lambda = g^2_{YM} N_c$ being  the 't
Hooft coupling~\footnote{This strong coupling result was first obtained by a direct use of AdS/CFT in Ref. \cite{Brower:2006ea}. The fact that the intercept is lowered from $j=2$  was strongly advocated in \cite{Brower:2000rp,Brower:1999tm} and its $1/\sqrt\lambda$ dependence  was  anticipated in \cite{Polchinski:2002jw}. This can also be arrived at by using a direct extrapolation from the weak coupling perturbative sum to the strong coupling limit, as done by Kotikov, Lipatov,  Onishchenko,  and Velizhanin~\cite{Kotikov}. See also \cite{Stasto:2007uv} and additional discussion in Sec. \ref{sec:DGLAP}.}.  It is useful to compare the dependence of the Pomeron intercept on the 't
Hooft coupling at  weak and strong couplings. It is worth noting
that the BFKL Pomeron intercept for ${\cal N}=4$ SYM, calculated to
second order, meets with the BPST intercept near $\alpha_{P}\simeq 1.2
\sim 1.3$, as shown in Fig. \ref{fig:Interceptb}, precisely in the
range as extracted from the HERA data. At these values, the 't Hooft
coupling is relatively large, around $\lambda \simeq 10$. For such a
large coupling value, it is important to explore the strong coupling
approach as a more efficient procedure for addressing physical
processes such as DIS. (There have already been several phenomenological studies of AdS/CFT which consider $\lambda$ in this range. For example in \cite{Kovchegov:2009yj}, strong coupling is
defined as $\lambda \geq 5$.)

AdS/CFT, or gauge/string duality, in principle allows a description of
conformal theory at  strong coupling by a weak coupling gravity
dual in an AdS background. However, QCD can be considered conformal
only approximately at best.  A conformal theory can never fully
reproduce all experimental results due to the lack of a scale and the
absence of confinement. However, at $Q^{2}$ sufficiently large,
partons inside the proton are expected to be free, and a conformally
invariant description could be a good approximation. Conversely, at
smaller $Q^2$ values, it is reasonable to expect that confinement
effects should be felt. Equally important is the phenomenon of
``saturation'', which should become important due to higher order
Pomeron-exchanges. In AdS/CFT, these non-linear effects come from
eikonalization. In contrast, in weak coupling, saturation has been
addressed primarily by considering non-linear evolutions such as the
BK equation~\cite{Balitsky:1995ub,Kovchegov:1999ua,Kovchegov:1999yj}.

From AdS/CFT, high energy scattering can be visualized as taking place in a 3-dimensional transverse Euclidean $AdS_3$, i.e., in addition  to the usual 2-dimensional impact-space $\vec b$, there is also the $AdS$-radial direction $z$. In \cite{Brower:2007qh, Brower:2007xg}, it has been shown, for two-to-two scattering involving on-shell hadrons, the amplitude in an eikonal sum can be expressed as
\begin{equation}
A(s,t)=2i s\int d^{2}b e^{i \vec q \cdot \vec b} \int dzdz'P_{13}(z)P_{24}(z')\big\{ 1- e^{i \chi(s,b,z,z')}\big\} ,\label{eq: scattering amplitude}
\end{equation}
where $P_{ij}(z)= \sqrt{-g(z)} (z/R)^2 \phi_i(z) \phi_j(z) $  involves a product of two external normalizable wave functions, satisfying normalization condition: $\int dz \sqrt{-g(z)} (z/R)^2 \phi_a(z) \phi_b(z) =  \delta_{ab}$.
The eikonal,
$\chi$,
is related to a BPST Pomeron kernel \cite{Brower:2006ea, Brower:2007qh, Brower:2007xg}
\begin{equation}
\chi( s,b,z,z')=\frac{g_{0}^{2}}{2{s}}(\frac{R^{2}}{zz'})^2K({s},b,z,z').   \label{eq:kernel2eikonal}
\end{equation}
 In the conformal limit, the Pomeron kernel can be expressed in a simple closed form and the eikonal becomes a function of two conformal invariants, a longitudinal boost $\tau$ and a variable $\xi$ relating to the chordal distance  in transverse $AdS_3$,
\bea
\tau &= & \log (\alpha' \hat s ) =\log({\rho}zz's/2)\; , \label{eq:tau} \\
 \xi &=&\sinh^{-1}\left( \frac{b^{2}+(z-z')^{2}}{2zz'}\right ).\label{eq:xi}
\eea
 where $\rho\equiv {2}/\sqrt{\lambda}=2\alpha'/R^2$ and  $\hat s=\frac{zz'}{R^{2}} s$.
It is also possible to discuss confinement deformations, leading to a unified treatment for the  Pomeron physics from UV to IR.   For definiteness, as done in \cite{Brower:2006ea, Brower:2007qh}, effects of confinement can be illustrated analytically by adopting  a model with a hard-wall cutoff.

We extend in this paper   the above treatment  to the case involving  two external currents, appropriate for DIS~\cite{Polchinski:2002jw}.  The  structure functions in DIS can be extracted from the off-shell $\gamma^*p$ total cross sections, which, from optical theorem, are proportional to the imaginary part of the forward off-shell $\gamma^*p$ amplitudes,  ${\rm Im} A(s,0)$. We therefore need to consider
the states of an offshell photon. This will require a replacement in $P_{13}$ by  wave functions divergent on the AdS boundary. As is well known, in
closed string theory there is no photon state. In \cite{Polchinski:2002jw} Polchinski
and Strassler have considered a $R-$boson propagating through the bulk
that couples to leptons on the boundary. To be specific, for $F_2(x,Q)$, which corresponds to  the sum of the transverse and longitudinal cross sections, the appropriate $R-$boson wave functions which mimic the vector currents  involve both  $K_0(Qz)$ and $K_1(Qz)$. After removing the associated polarization vectors and one factor of fine structure constant $\alpha_{em}$, one arrives at
\begin{equation}
P_{13}(z)\rightarrow P_{13}(z,Q^2)=\frac{1}{z}(Qz)^{2}(K_{0}^{2}(Qz)+K_{1}^{2}(Qz)), \label{eq:currentF2}
\end{equation}
and, for $x$ small where the Pomeron dynamics is applicable,
\be
F_2(x,Q^2) = \frac{ Q^2}{ 2\pi^2} \int d^2b \int dz \int dz' P_{13}(z,Q^2) P_{24}(z')   {\rm Re}\left ( 1-e^{i\chi( s,b,z,z')}\right)  \label{eq:DISeikonal}
\ee
(See also \cite{Hatta:2007cs,Kopeliovich:2009yw}). A similar replacement would lead to the second structure function, $F_1(x,Q^2)$,
\be
2x F_1(x,Q^2) = \frac{ Q^2}{ 2\pi^2} \int d^2b \int dz \int dz' P_{13}(z,Q^2) P_{24}(z')   {\rm Re}\left ( 1-e^{i\chi( s,b,z,z')}\right)  \label{eq:F1}
\ee
with
\begin{equation}
P_{13}(z)\rightarrow P_{13}(z,Q^2)=\frac{1}{z}(Qz)^{2}K_{1}^{2}(Qz).  \label{eq:currentF1}
\end{equation}
  We shall next examine the usefulness of this strong coupling formalism by testing  against the recently published combined H1-ZEUS data set. To simplify the discussion, we will focus primarily on $F_2$.

\section{DIS at Strong Coupling and the BPST Kernel}\label{sec:BPSTformalism}

Our goal is to calculate the structure function $F_{2},$ which is
given in terms of the cross section by
\begin{equation}
F_{2}(x,Q^{2})=\frac{Q^{2}}{4\pi^{2}\alpha_{em}}(\sigma_{T}+\sigma_{L}),\label{eq:f2 cross section}
\end{equation}
 where $\sigma_{T}$ and $\sigma_{L}$ correspond to the transverse
and longitudinal cross sections respectively. From the Optical Theorem, one can express the cross section in terms of the imaginary part of the forward scattering
amplitude, i.e., $\sigma_{total} = (1/s) Im A(s,t=0)$.  In principle Eq. (\ref{eq: scattering amplitude}) could be used for studying any scattering process where the Pomeron exchange
approximation is appropriate, by plugging in the appropriate wavefunctions
for the states $P_{13}(z)$ and $P_{24}(z')$. For our present application,
we consider deep inelastic scattering and the appropriate $R-$boson wave functions for $F_2$  can be taken to be  that given by (\ref{eq:currentF2}).     Note that, strictly speaking, the use of (\ref{eq:currentF2}) is valid only for $Q^2$ large and near the $AdS$ boundary where confinement effects can be ignored.

 Treating  the BPST Pomeron exchange to first order, we can then express the structure function $F_2$  in terms of the eikonal linearly as
 \begin{equation}
F_2(x,Q^2) =(Q^2/2\pi^2)\int d^{2}b\int dzdz'P_{13}(z,Q^2)P_{24}(z')Im\; \chi( s,b,z,z').\label{eq:forwardscatteringamplitude}
\end{equation}
with $P_{13}(z,Q^2)$ given   by  Eq. (\ref{eq:currentF2}).   We see that for $Qz>1$, $P_{13}(z,Q^2)$  rapidly decays to zero because of the Bessel
functions $K_0$ and $K_1$.  For small $Qz$, with $P_{13}(z,Q)$ bounded and, as we shall show shortly, $\int d^{2}b Im\; \chi( s,b,z,z')$ vanishing faster than $0(z^2)$,  it follows that the integrand  is peaked around $z\sim\nicefrac{1}{Q}.$   We also note that, after a simple calculation, $\int_0^\infty dz (zQ)^2 P_{13}(z,Q^2)=1.$

\subsection{DIS in the  Conformal Limit}
Let us begin by first considering the conformal limit. In this limit, the imaginary part of the BPST Pomeron kernel, $ Im K (s,b,z,z')$  can be expressed in a closed form, \cite{Brower:2006ea}, and, from Eq. (\ref{eq:kernel2eikonal}), it leads to
\begin{equation}
Im\; \chi( s,b,z,z') =   \frac{g_{0}^{2} }{16\pi }\sqrt{\frac{\rho }{\pi}}e^{(1-\rho)\tau}\frac{\xi}{\sinh \xi} \frac{\exp(\frac{-\xi^{2}}{\rho\tau})}{\tau^{\nicefrac{3}{2}}},\label{eq:ImK conformal}
\end{equation}
 where
$\tau$ and $\xi$ are given by (\ref{eq:tau}) and (\ref{eq:xi}) respectively.
The real part, $Re\; \chi(s,b,z,z')$, can be reconstructed via a dispersion integral, or, equivalently, by solving a derivative dispersion relation. As noted earlier, due to conformal invariance,  $Im\chi(s,b,z,z')$ depends only on two variables,  $\tau$, (or $\hat s$),  and $\xi$. The $b$-space integration of this kernel  can be carried out explicitly~\cite{Brower:2007xg},  leading to
\bea
\int d^{2}b\;  Im\;  \chi (s,b,z,z')
&=&
 \frac{g_{0}^{2} }{16}{\sqrt{\frac{\rho^3}{\pi } } } \; (zz' )\;  e^{(1-\rho) \tau}\frac{\exp(\frac{-(\log z -\log z')^2}{\rho\tau})}{\tau^{\nicefrac{1}{2}}}.  \label{eq:integratedConformalEikonal}
\eea
 This  expression  now takes on the familiar form of ``diffusion" in $|\log z|$. This was first noted in Ref. \cite{Brower:2006ea}.

Let us next turn to the factor $P_{24}$. For the proton, one is dealing with a normalizable state. Strictly speaking, this necessarily breaks the conformal limit, e.g., to
model QCD we need to  introduce a cutoff at $z_{0}\sim1/\Lambda,$ where
$\Lambda$ is the confinement scale. This construct would be sufficient for describing glueballs.  However, for baryons, one needs to go beyond the standard confining deformation and additional holographic prescriptions will have to be adopted~\cite{Kopeliovich:2009yw,Levin:2009kk,Domokos:2009hm}.  It suffices to assume for now that the $AdS$ wave-function for the proton is localized near the
cutoff in the bulk,  $z_{0}$.

Putting these two together, we have, for DIS in the strong coupling conformal limit, that  the structure function $F_2$ can be expressed as
\be
F_2(x, Q^2)= \frac{g_{0}^{2} \rho^{\nicefrac{3}{2}}}{32 \pi^{\nicefrac{5}{2}}}     \int dzdz'P_{13}(z,Q^2)P_{24}(z') {(zz' Q^2)}\; e^{(1-\rho)\tau}\frac{\exp(\frac{-(\log z -\log z')^2}{\rho\tau})}{\tau^{\nicefrac{1}{2}}}\label{eq:f2conformal}
\ee
where $x\simeq Q^2/s$ and $P_{13}$ given by (\ref{eq:currentF2}).
Note that $F_2$ is uniquely determined once  the  parameter $\rho$ and the overall strength factor $g_0^2$ are specified. There are two important features for Eq. (\ref{eq:f2conformal}). It is easy to verify, as noted earlier, that the integrand vanishes at least as fast as  $z^{1-\rho} $ at $z=0$ and is exponentially damped for $z>>1/Q$. It follows that the dominant contribution to the integral coming from $z=O(1/Q)$. With $z'=O(1/\Lambda_{qcd})$,  the  factor $e^{(1-\rho)\tau}$ leads to
\be
e^{(1-\rho)\tau} \sim (1/x)^{1-\rho}\;,  \label{eq:Reggegrowth1}
\ee
with $Q^2$ fixed. The limit $x\rightarrow0$ will lead to a fast rising in $F_2$, violating the standard Froissart bound. In terms of the eikonal $\chi(\hat s,b,z,z')$, it is generally accepted that the condition $\chi = O(1)$ would signal the onset of ``saturation''. Note that this is a ``local condition'' in the three-dimensional transverse space, $z$ and $\vec b$, with $z'$ fixed.   One of the main goal of our current analysis is to investigate the importance of  saturation in the HERA range.  We will come back to this discussion in Sec. \ref{sec:Saturation}.

The second key feature for (\ref{eq:f2conformal}) is the last exponential factor,
\be
\exp\left(\frac{-(\log z -\log z')^2}{\rho\tau}\right)
\label{eq:diffusion}
\ee
With  $z\sim 1/Q$, this corresponds to diffusion in $\log Q $, analogous to diffusion in ``virtuality'', found in weak coupling BFKL  dynamics.  As we shall see shortly, we find this diffusion effect playing a crucial role in understanding the $Q^2$ dependence for the $\epsilon_{eff}(Q^2)$ observed at HERA, as exhibited in Fig. \ref{fig:Intercepta}.

\subsection{Regge and DGLAP Connection}\label{sec:DGLAP}

Although diffusion is common to both the weak and the strong couplings, the diffusion coefficient  in the strong coupling, (\ref{eq:diffusion}), is correlated with the Regge growth, (\ref{eq:Reggegrowth1}), leading to $Q^2$-independence for $\int_0^1 dx F_2(x,Q^2)$  in the large $Q^2$ limit, i.e.,
\be
\frac{d \;\;\;\;}{dQ^2} \int_0^1 dx F_2(x,Q^2)  = 0.\label{eq:anomalousDim}
\ee
It should be stressed that this feature is not shared by the conventional weak coupling BFKL approach.   Eq. (\ref{eq:anomalousDim}) is required by energy-momentum conservation, or, more technically, the vanishing of anomalous dimension for the second moment for the structure functions.
 Indeed,  the signature feature of the BPST treatment is based on perturbing about the supergravity limit where the energy-momentum conservation is guaranteed. Consider the  moments for the structure function $F_2$,
 \be
 M_n(Q^2) = \int dx x^{n-2} F_2(x,Q^2)   \label{eq:moments}
 \ee
 From OPE, one has, for $Q$ large,
 \be
 M_n(Q^2)\rightarrow ( Q)^{-\gamma_n}
 \ee
 where $\gamma_n$ is the anomalous dimension for the twist-two operators, appropriate  for the DIS.

 From our strong coupling analysis, these anomalous dimensions can be read off the more general ``dimension-spin'' curve, Fig. \ref{fig:deltaJ}. At  strong coupling, $\lambda>>1$,  one finds  to leading order in $\rho=2/\sqrt\lambda$,  the relation between scaling dimension $\Delta$ and spin $j$ is parabolic,
 \be
j= 2 + \rho [(\Delta-2)^2 /2 -1 ]
 \ee
As explained in \cite{Brower:2006ea}, this follows  from the physical state condition, $L_0=\bar L_0=1$, for string theory at strong coupling. The DGLAP dimensions and the Pomeron intercept are determined by the same on-shell vertex operator.   The symmetry about $\Delta=2$ is due to conformal invariance, and, at $j=2$, one has $\Delta=4$, due to energy-momentum conservation. Although we have drawn in  Fig. \ref{fig:deltaJ} for weak coupling with the curve maintaining the symmetry about $\Delta =2$ and passing through $\Delta=4$ and $j=2$, this feature is absent in a conventional perturbative treatment \cite{Kotikov}. 

In evaluating $M_n$, Eq. (\ref{eq:moments})  acts as a $J$-plane Mellin transform in $s$, thus allowing  one to identify  the moment $n$ with the spin $j$. A straight forward calculation then leads to
 \be
 \gamma_n = 2\sqrt{1+ (n-2)/\rho} -n  \label{eq:anomalous}
 \ee
At $n=2$, $\gamma_2$ vanishes, as expected.\footnote{Note that, by focusing on small-$x$, our calculation for $\gamma_n$ becomes unreliable for $n$ large, which reflects strongly the behavior of structure functions near $x=1$.} Because of our better control of the anomalous dimensions in the strong coupling at $n=2$, our BPST Pomeron provides a dual approach  to the DGLAP treatment, so long as a linear evolution is valid at the HERA energy.  Instead of evolving in $Q^2$, the $x$-evolution is controlled by the minimal of the $\Delta-J$ curve at $j_0=2-2/\sqrt \lambda=2-\rho$, our BPST intercept.

\vskip .5in
\begin{figure}[bthp]
\begin{center}
\includegraphics[height=0.4 \textwidth,width=0.5\textwidth]{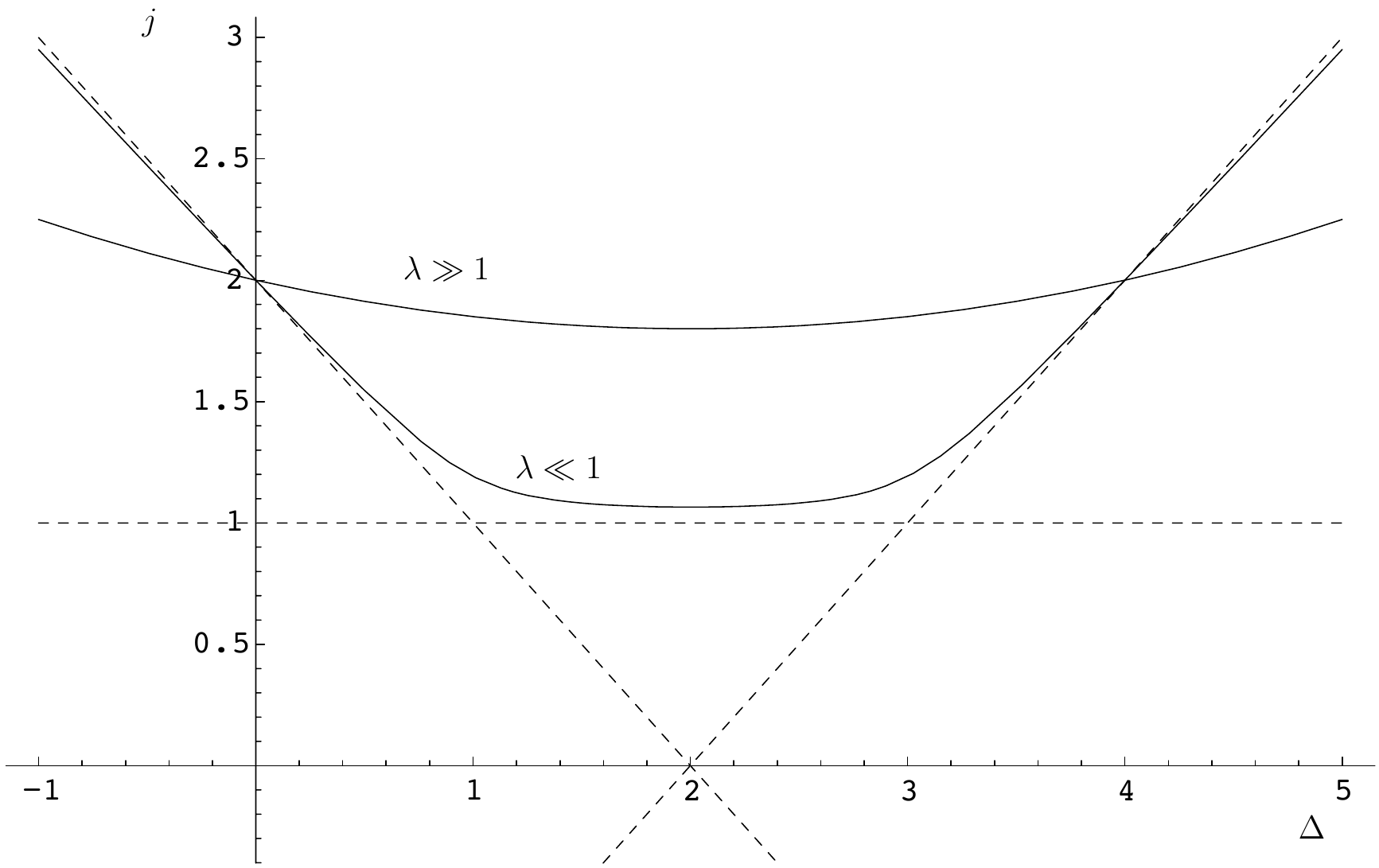}
\caption{ $\Delta-J$ curve.   This figure has been reproduced from Fig. 2  of \cite{Brower:2006ea}.}
\label{fig:deltaJ}
\end{center}
\end{figure}

\subsection{Confinement}\label{sec:confinement}
Let us next turn to the case with confinement, using the hard-wall model for illustration. In this model, we introduce a sharp boundary in the $AdS$
$z$ coordinate at some value $z_{0}.$ This leads to a mass gap and
a discrete spectrum of states. Our starting points are still Eqs.
(\ref{eq:f2 cross section}-\ref{eq:forwardscatteringamplitude}), but  a modified Pomeron kernel has to be used. Such a  kernel can formally be represented in a spectral representation, but it cannot be expressed in a  simple  closed form as the conformal case, i.e., (\ref{eq:ImK conformal}).  Fortunately, for a single Pomeron contribution, we only need to  evaluate the total contribution after integration over the impact parameter $\vec b$. That is, if we go to momentum space,
\be
\chi(s,t,z,z')= \int d^2b e^{iq_\perp b_\perp} \chi(s,b,z,z'),  \label{eq:chi-t}
\ee
 where $t=-q_\perp^2$, we only need to know $\chi(s,t,z,z')$ evaluating in momentum space at $t=0$. For this, a closed form has been found for the BPST hard-wall kernel, \cite{Brower:2007xg}, leading to
\begin{equation}
Im \; \chi_{hw}(s,t=0,z,z')=Im\; \chi_c(\tau,0,z,z')+{\cal F}(z,z',\tau)\; Im\; \chi_c(\tau,0,z,z_0^2/z'),\label{eq:ImK hard-wall}
\end{equation}
 where  $Im\; \chi_c(\tau,0,z,z')$ is the conformal expression given by the right hand side of Eq. (\ref{eq:integratedConformalEikonal}), and
\begin{equation}
{\cal F}(z,z',\tau)=1-2\sqrt{\rho\pi\tau}e^{\eta^{2}}erfc(\eta),\,\,\,\,\,\,\,\,\,\,\,\,\eta=\frac{-\log\frac{zz'}{z_{0}^{2}}+\rho\tau}{\sqrt{\rho\tau}}.\label{eq:F, eta}
\end{equation}
It follows that, with confinement,
\be
F_2(x, Q^2)=\frac{g_{0}^{2} \rho^{\nicefrac{3}{2}}  }{32 \pi^{\nicefrac{5}{2}}} \int dzdz'P_{13}(z,Q^2)P_{24}(z') {(zz' Q^2)}\; e^{(1-\rho)\tau}\left(\frac{e^{-\frac{\log^{2}\nicefrac{z}{z'}}{\rho\tau}}}{\tau^{\nicefrac{1}{2}}}+{\cal F}(z,z',\tau)\frac{e^{-\frac{\log^{2}\nicefrac{zz'}{z_{0}^{2}}}{\rho\tau}}}{\tau^{\nicefrac{1}{2}}}\right)\label{eq:f2hw}
\ee

\begin{figure}[bthp]
\begin{center}

\includegraphics[width=1\textwidth]{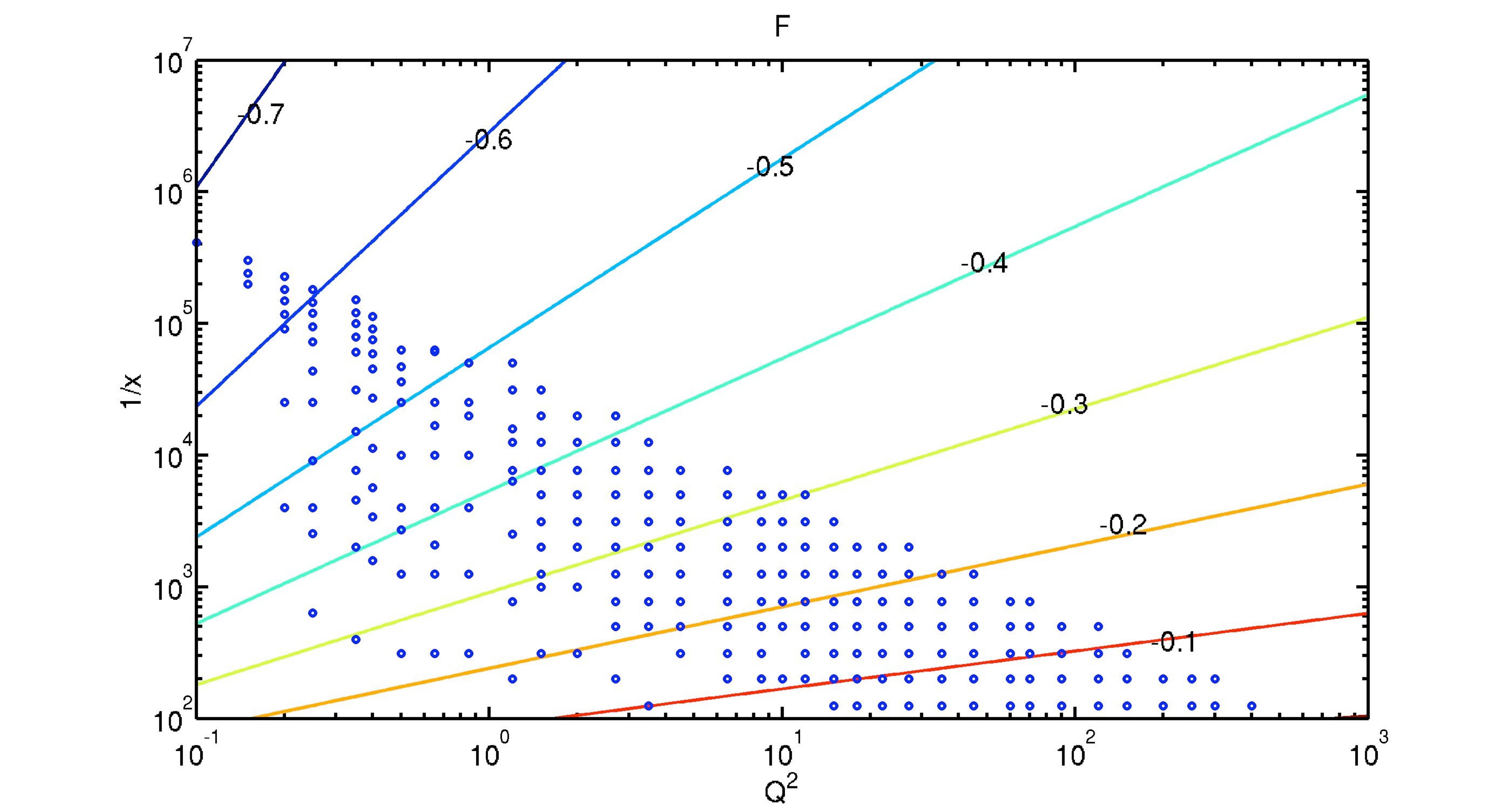}

\caption{ Contour plot for coefficient function ${\cal F}$ as a function of $log (1/z)$ and $\log (1/x)$, with $z'\simeq z_0$ fixed, $z_0\sim \Lambda_{QCD}^{-1}$. Confinement effects become important in the region where $|{\cal F}|  > 0.3 \sim 0.5 $. We have also shown the entire combined H1-ZEUS small-$x$ data points ~\cite{:2009wt} as dots by identifying $Q=1/z$. }
\label{fig:F-factor}
\end{center}
\end{figure}

The
first term on the right-hand side of  (\ref{eq:f2hw}) is precisely the same as that in the conformal limit,
(\ref{eq:integratedConformalEikonal}) and is model independent. The second term, which can be expressed as a linear superposition of contributions from a set of image charges,  is dependent upon
the details of how the space is cutoff at $z_{0}$ and hence model
dependent. Nevertheless it is reasonable to assume its presence can provide a qualitatively reliable estimate of the effects  of confinement.
The coefficient function, ${\cal F}(z,z',\tau)$, (\ref{eq:F, eta}),  at fixed $z,z'$, goes to $1$ as $\tau\rightarrow0$ and to $-1$ as
$\tau\rightarrow\infty.$ Hence, at small $x$, i.e., $\tau$ large, ${\cal F}\rightarrow -1$ and  confinement
leads to  a partial cancelation for the growth rate relative to that of  the conformal BPST Pomeron. Moreover,
since ${\cal F}$ is continuous, there will be a region over which ${\cal F} \sim 0$, and, in this region, there is little difference between the hard-wall and  the conformal results.

In Fig. \ref{fig:F-factor}, we provide a contour plot for ${\cal F} $ as a function of $\ln x$ and $\ln z$, with $z'$ fixed near $z_0$. Using the fact that, from $P_{13}(z,Q^2)$ is peaked at  $z\simeq 1/Q$, this can also be interpreted as a contour plot in $1/x$ and $ Q^2$, with $\ln Q \simeq -\ln z$. Anticipating our subsequent fit which fixes the scale with $z_0\sim \Lambda_{qcd}$, we have also exhibited   the entire set of the recently combined ZEUS-H1 small-$x$ data points  as open dots on this plot.  We note  that, over a significant region, e.g., $Q^2> O(1) \; GeV^2$,  $|{\cal F}| $ is small and  the conformal kernel remains a reasonable approximation. Equally important is the observation that the region where confinement effects can be significant, roughly defined by the condition $|{\cal F}| > c$, $c\simeq 0.3-0.5$, goes beyond the narrow region of $Q^2 =O (1) \; GeV^2$. The line for $c=0.5$ is also drawn in Fig.  \ref{fig:HERA_LHC} indicating schematically the transition into confinement region. This observation goes against the conventional belief that confinement effects can be ignored for $Q^2$  large, independent of $x$. This also casts doubt on the assumption that saturation at $Q^2$ large can be understood without taking confinement into account.

 Eq. (\ref{eq:f2hw}) is the expression which we will use shortly for comparing
with experimental results. This expression retains the two key features noted earlier for the conformal result, namely, the presence of the dominant term in $\tau$, leading to
$
(1/x)^{1-\rho}
$
rise, and the presence of diffusion in $z$. It can also be shown that (\ref{eq:anomalousDim}) still holds in the large $Q^2$ limit.

\section {Confronting the HERA Data -- Linear Treatment} \label{sec:linearfit}

We now  carry  out a test of our strong coupling DIS analysis  by fitting the recently published combined H1-ZEUS data set. We will first test    single BPST Pomeron models, both conformal and with confinement, before trying the eikonal form in the next section.  We restrict ourselves to the small-$x$ region, i.e., $x<0.01$. This remains a  relatively large  data set with 249 points, and in particular, the set  extends to a large range of  $Q^2$ values, from  $0.1\, GeV^{2}$ to $400\, GeV^{2}$. We  find that the confinement-improved  treatment allows a surprisingly good fit to all HERA small-$x$ data.     In contrast,  the fit based on the conformal  BPST Pomeron breaks down  in the low-$Q^2$ region.

Before proceeding further, we must specify more precisely the $AdS$ wave function, $\phi_p(z)$, associated with the proton. Unfortunately, other than the expectation that it be  normalizable, i.e., $\int dz \sqrt {-g(z)} (z/R)^2 |\phi_p(z)|^2=1$, one cannot determine $\phi_p$ without an explicit strong coupling model for baryon. For the current analysis, we will assume that the wave function is sharply peaked near the IR boundary $z_0$, with $1/Q'\leq z_0$, with $Q'$ of the order of the proton mass~\cite{Kopeliovich:2009yw,Levin:2009kk,Domokos:2009hm}. For simplicity,
we will simply  replace $P_{24}$ by a sharp delta-function
\begin{equation}
P_{24}(z')\approx \delta({z'}-1/Q').
\label{eq:p24}
\end{equation}
Similarly, as noted earlier, the integral over $P_{13}(z,Q^2)$ in (\ref{eq:f2conformal}), and also in (\ref{eq:f2hw}), is centered at $z=1/Q$, with  $\int dz (zQ)^2 P_{13}(z,Q^2)=1$.  As emphasized earlier, (\ref{eq:f2conformal}) follows only for $Q^2$ large, and there will be modifications coming from confinement for $Q^2$ small. To simplify the discussion below, we will again make the local approximation by replacing $P_{13}(z,Q^2)$ by a delta-function
\begin{equation}
P_{13}(z)\approx C \delta(z-1/Q), \label{eq:p13}
\end{equation}
with $C\simeq 1$. These constitute our model-dependent  parametrizations.
 With these substitutions,
the $z$ and $z'$ integrations can be performed trivially, leading to $F_2$ given essentially by the Pomeron kernel. It follows  that the structure function, (\ref{eq:f2conformal}),  becomes
\begin{equation}
F_{2}(x,Q^2) =\frac{g_{0}^{2} }{32 \pi^{\nicefrac{5}{2}}}\rho^{\nicefrac{3}{2}}\frac{Q}{{Q'}}e^{(1-\rho)\tau}\; \left(\frac{\exp (-\frac{\log^{2}(\nicefrac{Q}{{Q'}})}{\rho\tau})}{\sqrt{\tau }}\right).\label{eq:f2conformalb}
\end{equation}
with  $\tau$ as a function of $x$ and $Q^2$: $\tau(x,Q^2)=\log[(\frac{\rho Q }{2 Q'})(\frac{1}{ x})]$.
The hard-wall single-Pomeron contribution, (\ref{eq:f2hw}), can similarly be simplified,  with $\eta$ in (\ref{eq:F, eta}) expressed as $\eta(x,Q^2)=\frac{\log(z^2_{0}Q {Q'})+\rho \tau(x,Q^2)}{\sqrt{\rho \tau(x,Q^2)}}.$

 To see how well our strong coupling expression comes  close to the
 experimental data,  we begin by first restricting  ourselves to a smaller set of ZEUS data with $x < 0.01$ and a
 range of $Q^{2}$ from $0.65$ to $650\, GeV^{2}.$ Although there are
 data at lower values of $Q^{2},$ we do not consider them initially
 since we would like to avoid the region where confinement effect can
 be expected to be important.  We will also begin by first testing
 against the previously published ZEUS data~\cite{Breitweg:1998dz,Chekanov:2001qu}, so we can compare the
 goodness of our fit relative to other published fits done by others
 without invoking AdS/CFT~\cite{Soyez:2004ni,Block:2006dz,Berger:2007vf}.  Another reason for avoiding the small
 $Q^2$ region is because in our formalism it is natural to restrict
 $Q>Q'$, where $1/Q'\simeq z'$ characterizes the ``size'' of the
 proton and, as we shall see, will turn out to be $\approx0.5\, GeV.$

 We fit the $F_2$ to the ZEUS data using Matlab with $4$ free
 parameters, which we choose to be : $\rho$, $ g_0$, $z_{0}$, and
 $ Q'$.  We
 have carried out fits for both the conformal and the hard-wall
 models.  In total we had $160$ different data points with $32$
 different values of $Q^{2}.$   Surprisingly, good fits can be achieved for both.  To be specific, we
 obtain the ``best fit'' for the confining hard-wall model, with the
 following values: $\rho=0.7716\pm 0.0103, \; g_0^2 =106.01\pm 3.10, \; z_{0}=6.60\pm 1.50 \, GeV^{-1},
 \;  Q' =0.5322\pm 0.0465\; GeV \, $.  These values are ``reasonable'', within our
 general expectations. For instance, the value of $\rho\simeq 0.77$
 lies within the transition region between strong and weak
 coupling. The value of $z_0$ is also consistent with our expectation
 of $z_0\sim O( \Lambda_{qcd}^{-1})$.  With $z_0$ fixed, the value for $Q' $ is again reasonable.
 The chi-square value per
 degree of freedom \footnote{We calculated the $\chi_{d.o.f.}^{2}$
   using a standard formula,
   $\chi_{d.o.f.}^{2}=\frac{1}{N-p}\sum_{i}\frac{(O_{i}-E_{i})^{2}}{\sigma_{i}^{2}},$
   where $N$ is the number of points, $p$ is the number of parameters,
   and $O_{i},E_{i}$ and $\sigma_{i}$ are the observed, expected, and
   the total error of the observed values respectively. (See, for instance, P. Fornasini, ``The Uncertainty in Physical Measurements'', Springer 2008.)  We did not
   eliminate any points in our analysis. }  we calculate for our best
 fit is $\chi_{d.o.f.}^{2}=0.69.$ An equally good fit can also be
 obtained using the conformal Pomeron, with best fit values:
 $\rho=0.774\pm 0.0103, \; g_0^2 =110.13\pm 1.93 , \; Q' = 0.5575\pm 0.0432 \; GeV $, and the
 corresponding chi-square value is $\chi_{d.o.f.}^{2}=0.75.$

Armed with this success, we next carried out an expanded study for the recently published combined H1-ZEUS data set~\cite{:2009wt}, keeping only small-$x$ data. With $x<10^{-2}$,  the set now extends to much smaller $Q^2$ values, with  $Q^2$ ranging from  $0.1\, GeV^{2}$ to $400\, GeV^{2}$, taking on 34 different $Q^2$ values. This is a  larger data set than the ZEUS set considered above, increasing from 160 data points to 249 points. These are the data points shown in Fig. (\ref{fig:F-factor}) as dots, with $z=Q^{-1}$.

With the inclusion of data at $Q^2 < 0.65 GeV^2$, we can no longer obtain acceptable fits by a single conformal BPST Pomeron. The best fit leads to an unacceptably large value of  $\chi^2_{d.o.f.}= 11.7$. The fit cannot be improved by the sieve-procedure~\cite{Block:2005qm}. That is, one cannot attribute the poor fit to the presence of "outliers".  In contrast, we  find that the confinement-improved  (hard-wall) treatment allows a surprisingly good fit to all HERA small-$x$ data, especially after applying the sieve-procedure.
We obtain the ``best fit'' for the confining hard-wall model, with the following  values
\begin{equation}
\rho=0.7792\pm 0.0034, \;   g_0^2 =103.14 \pm 1.68 , \; z_{0}=4.96 \pm 0.14\, GeV^{-1}, \; Q' =  0.4333\pm 0.0243 \; GeV  ,\,   \,\label{eq:parameters}
\end{equation}
These values differ from our earlier fit only slightly and they remain
reasonable, within our general expectations.  Without sieving, we
obtain $\chi_{d.o.f.}^{2}= 1.34$. With the elimination of 6 data
points as ``outliers'', each with $\chi^2>8$, the $\chi_{d.o.f.}^{2}$
improves to be $1.16$. The best cutoff is for $\chi^2>4$, leading to a
calculated chi-squared per degree of freedom $\chi_{d.o.f.}^{2}=1.07$
for our best fit given above~\footnote{With a $\chi^2$-cutoff at 8, we
  included a renormalization factor of ${\cal R}=1.0433$. When the  cut-off
  is at 4, one has ${\cal R}=1.2924$.}.

In Fig. \ref{fig:global} we show our fits to $F_2(x,Q^2)$ for all 34
different $Q^2$ as a function of $x^{-1}$. The fit for a single
conformal  BPST Pomeron is shown as blue curves and that for a single
hard-wall BPST is shown as red curves.  It is evident that, for $Q^2$
small, the conformal model does not fit the data well. However, for
$Q^2> 4 \sim 5 \; GeV^2$, where data points exist, the conformal and the
hard-wall models remain comparable.

\begin{figure}[bthp]
\begin{center}
\includegraphics[height=1.0 \textwidth,width=1.0\textwidth]{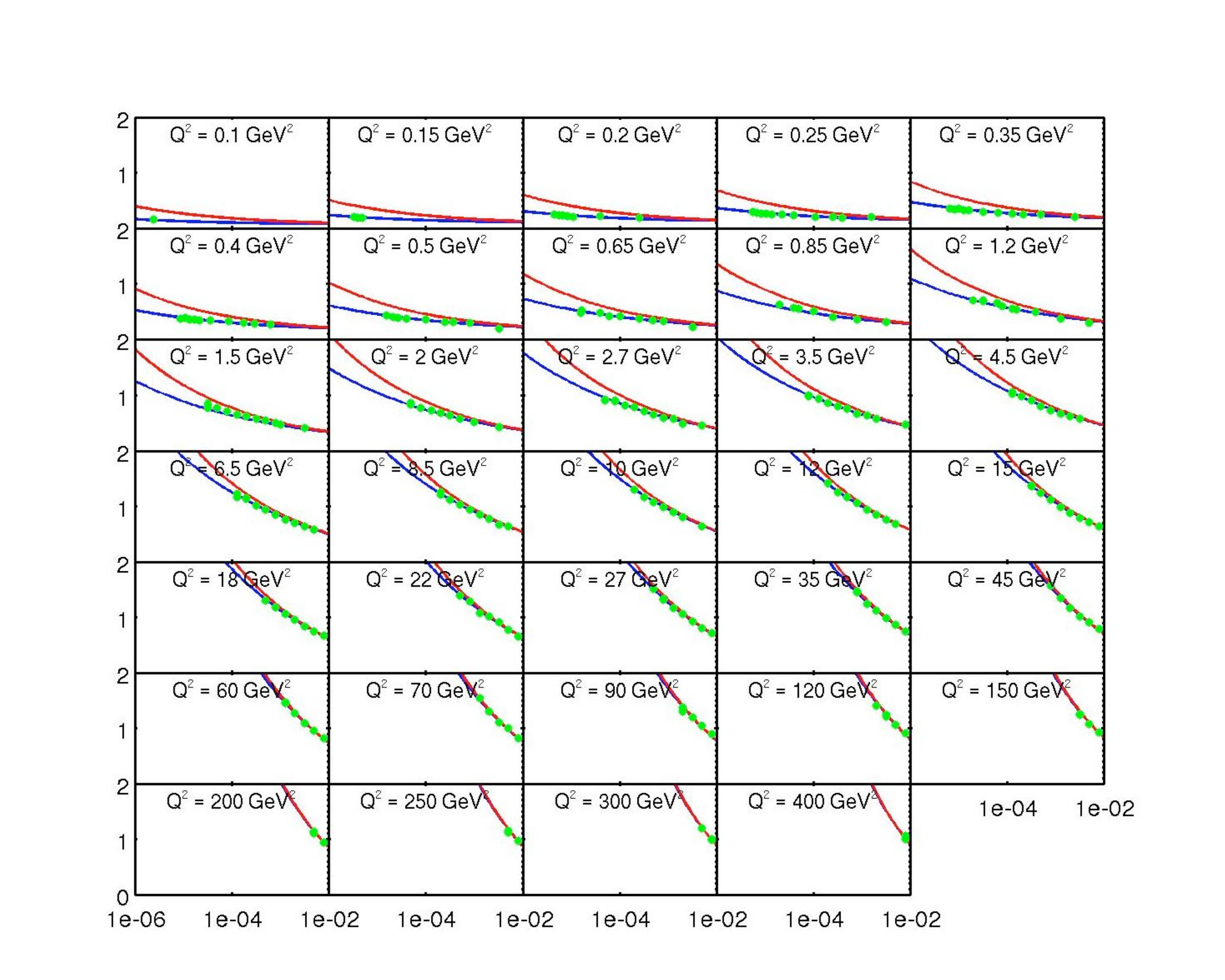}
\caption{ Global fits to the combined ZEUS-H1 small-$x$ data.  Dotted
 red lines are for single conformal BPST Pomeron and dotted blue   lines are for single hard-wall BPST Pomeron. }
\label{fig:global}
\end{center}
\end{figure}

We can further test our result by turning to the HERA effective
Pomeron intercept plot, Fig. \ref{fig:Intercepta}.  Intercept values  derived from
the conformal Pomeron are in red and those  from the hard-wall model are in blue. Much of the $Q^2$-dependence for the effective Pomeron
intercept can be attributed to the diffusion effect. As $Q^2$ becomes
large, diffusion is enhanced, leading to an increase to the effect
rise in $F_2$ as $1/x$ increases. Let us next examine the hard-wall
model, (\ref{eq:f2hw}), where we see the effect of confinement is
embodied in the second term. For $Q^2 >>1\; GeV^2$, the difference
between these two at HERA energy range is small. As mentioned earlier,
from Fig. \ref{fig:F-factor}, with $|{\cal F} | << 1$ mostly in this
region, it is not surprising that the red and the blue lines converge
as $Q^2$ increases. However, as $Q^2$ decreases, one finds $|{\cal
  F}|$ increases, and the effect of ``destructive interference'' is
enhanced. In particular, in the extreme limit of $z\sim z' \sim z_0$,
the leading order term from the denominator of the diffusion factor,
$1/\sqrt \tau$, largely cancel, leading effectively to a $\tau^{-3/2}$
suppression. This enhanced suppression contributes to a lowered value
for the effective Pomeron at small $Q^2$, which is reflected by the
red-curve in Fig. \ref{fig:global}. That is, using the hard-wall model
as a guide, confinement leads to an effective lowering of the Pomeron
intercept for scattering of two ``soft'' states, i.e., states with wave
functions centered near the IR wall. This is consistent with the view
of ``soft Pomeron'' dominance for hadronic cross sections in the near
forward region, with an effective intercept $\alpha_P\simeq 1.08$. Of
course, in the region, one also expects eikonalization would
eventually become important. We turn to this issue in the next Section.

It is also worth noting that, under our local approximation where $z\simeq 1/Q$, one has, for $1/x$ large, 
\be
R=\frac{F_L}{F_T}\simeq \frac{\int_0^\infty dx x^{3-\rho}|K_0(x)|^2}{\int_0^\infty dx x^{3-\rho}|K_1(x)|^2}=\frac{2-\rho}{4-\rho},
\ee
where $F_L=2xF_1$ and $F_T=F_2-2xF_!$. With $0<\rho<1$, $R$ ranges from $1/2$ to $1/3$, achieving the maximum  in the supergravity limit when $\rho\rightarrow 0$. This differs significantly from that expected for $x$ large due to Callan-Gross relation. This is consistent\footnote{In \cite{Cornalba:2010vk}, $R\simeq (1+\omega)/(3+\omega)$, with $\omega=1-\rho$.} with that found previously in Ref. \cite{Cornalba:2010vk}.

\section{Nonlinear Evolution  and Saturation}\label{sec:Saturation}

  At strong coupling nonlinearity enters through eikonalization. When eikonalization becomes important, it also signifies the onset of ``saturation''.    Instead of expanding   Eq. (\ref{eq:DISeikonal}) to first order in the eikonal, a fully non-perturbative treatment is now required.
Clearly, nonlinearity becomes important only when
\be
\left |\chi( s,b,z,z')\right | \geq  O(1).  \label{eq:saturation}
\ee
In general, both the real and the imaginary parts of the eikonal  will have to be taken into account.  Note that this is a local condition, in the three-dimensional transverse $(\vec b, z)$ space.  (For a more traditional weak coupling approach, see \cite{Gribov:1983,Balitsky:1995ub,Kovchegov:1999ua,Kovchegov:1999yj}. See also \cite{Hatta:2007cs}, \cite{Kovchegov:2010uk}, and references therein.) 

Using the conformal eikonal as a guide, it is easy to convince ourselves that, for the HERA range, it is adequate to treat Pomeron linearly and one is far away from saturation.   In Fig. \ref{fig:eikonal}a, we show  a contour plot for the conformal eikonal, using the parameters given in Eq. (\ref{eq:parameters}), at zero impact parameter, $\vec b=0$. We have also made use of the fact that $z\simeq 1/Q$, which allows one to label all the HERA data set on the same plot. Note that, for $\vec b=0$, where the eikonal is the largest, the maximal value in the HERA range remains less than $0.2\sim 0.3$. The conformal eikonal decreases with increasing $|\vec b|$ as a power, when $|\vec b|>> z$.    We have carried out an eikonal analysis using the conformal eikonal. Not surprisingly, there is no improvement in the quality of the fit. We again find with unacceptable chi-square per degree of freedom, due primarily to the inadequate fit  in the region of small $Q^2$.  In order to answer the question regarding the onset of saturation,  we next turn to an eikonal treatment for the hard-wall model.

\begin{figure}[bthp]
\begin{center}

\includegraphics[height=0.3 \textwidth,width=.45\textwidth]{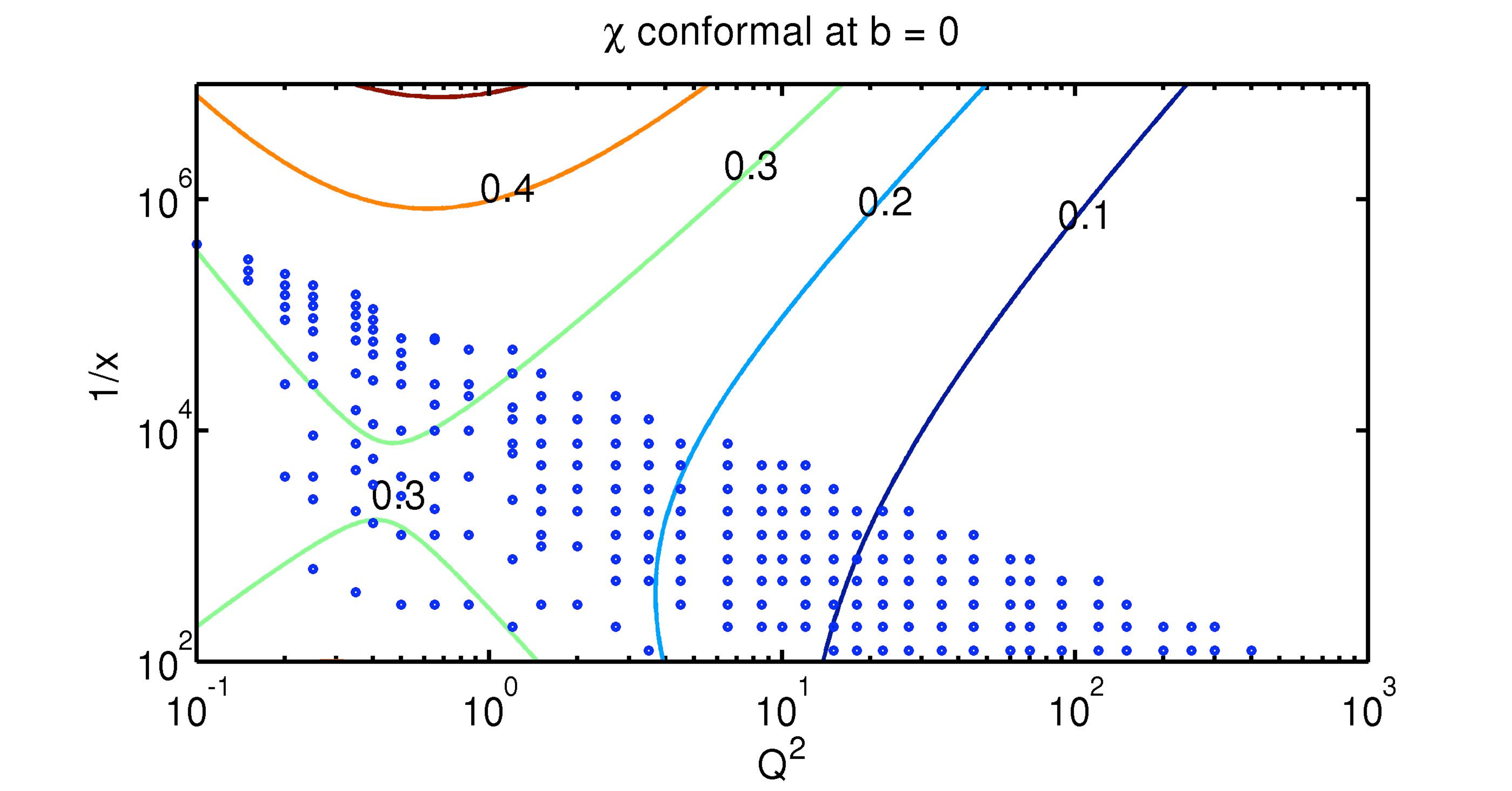}
\includegraphics[height=0.3 \textwidth,width=.45\textwidth]{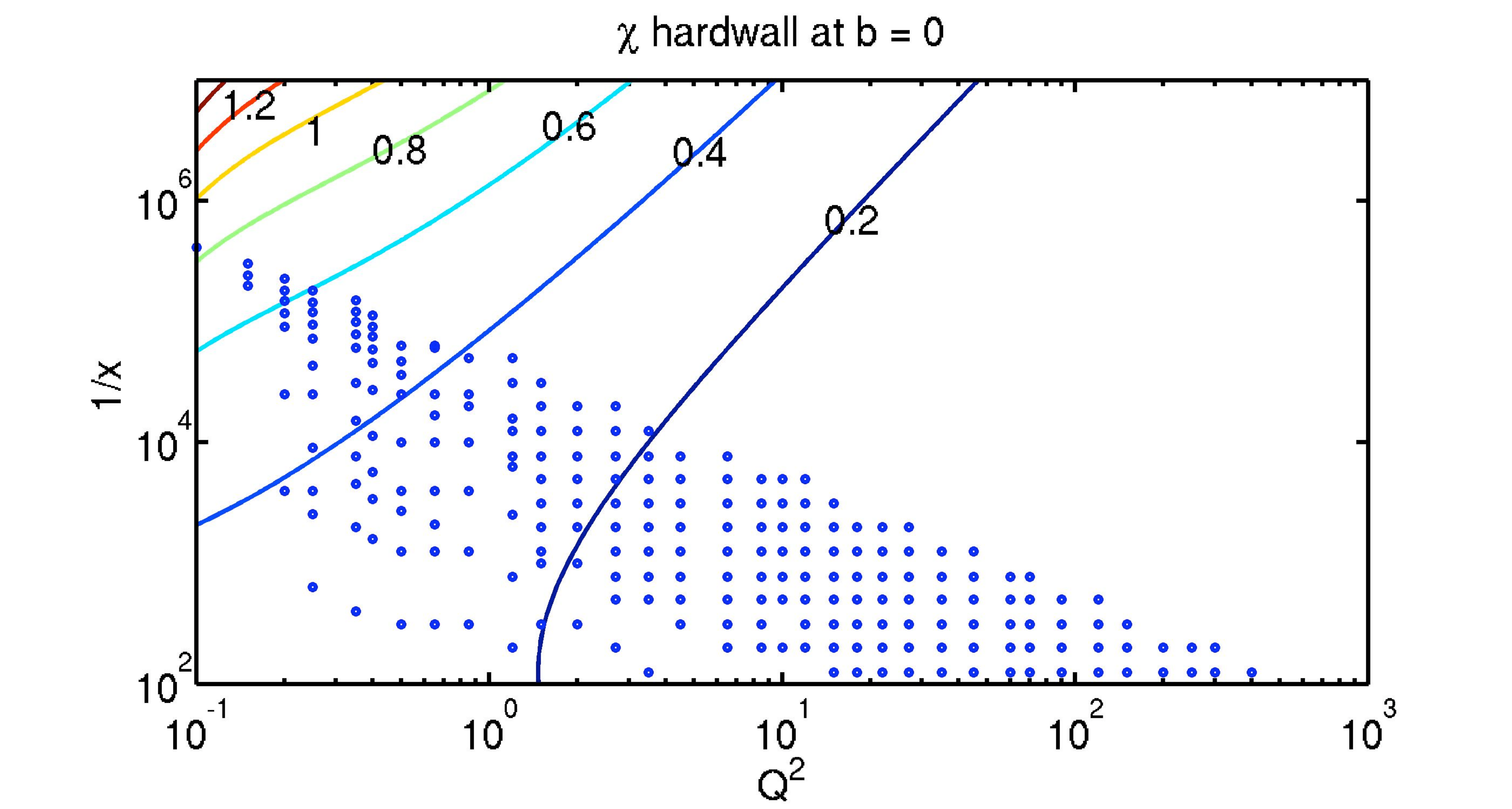}

\caption{ Contour for $Im\;\chi(s,b,z,z')$ for conformal and hard-wall respectively, with $z'\simeq z_0$ fixed, as a function of $x$ and $z$, ($z=Q^{-1}$), at zero impact parameter, $\vec b=0$. }
\label{fig:eikonal}
\end{center}
\end{figure}

The eikonal for the hard-wall model, $\chi_{hw}(s,b,z,z')$, can be obtained via a spectral representation \cite{Brower:2006ea, Brower:2007xg}. However, it is difficult  to cast $\chi_{hw}(s,b,z,z')$ in a form amenable for direct manipulation, except for its integrated form, $\int d^2b\; Im\; \chi_{hw}(s,b,z,z')$,  given by (\ref{eq:ImK hard-wall}).   Fortunately,  the effect of eikonalization at HERA is likely to be weak except for $Q^2$ small. We will therefore provide an approximate treatment   which incorporates the more important new features  due to confinement.

Let us begin by first working in a momentum representation, (\ref{eq:chi-t}) and focus on $Im\; \chi_{hw}(\tau,t,z,z')$.
 For $t\neq 0$, $Im\; \chi_{hw}(\tau,t,z,z')$ can be formally obtained by solving an integral equation
\bea
Im\; \chi_{hw}(\tau,t,z,z') &=& Im\; \chi_{hw}(\tau,0,z,z') \nn
&+&\frac{ \alpha_0t}{2}  \int_0^\tau d\tau'\int_0^{z_0} d \tilde z \; { \tilde z}^2\;  Im\; \chi_{hw}(\tau',0,z,\tilde z) Im\; \chi_{hw}(\tau-\tau' ,t, \tilde z,z') \label{eq:integralequation}
\eea
with   $Im\; \chi_{hw}(\tau,0,z,z')$ given by Eq. (\ref{eq:ImK hard-wall}). Solution to this integral equation can be investigated numerically, which is currently under way and will be reported separately. Here an approximate treatment will be provided.

The single-most important consequence of  confinement is the existence of a mass gap. In the conformal limit,  $ \chi_{c}(\tau,t,z,z') $ has a branch point  at $t=0$, which is responsible for a power-like fall off at large impact separation. For instance, in  the supergravity limit where $j\rightarrow 2$, this leads to the well-known cutoff  for $b$ large,
\be
\chi_c(\tau ,b,z,z')  \sim  {e^{\tau}}{b^{-6}}\;.
\ee
Incidentally, the eikonal becomes real in this limit. More generally,
a power decrease signals the presence of a $t=0$ singularity in the
momentum representation. In contrast, for hard-wall, because of
confinement, $\chi_c(\tau ,t,z,z')$ is regular at $t=0$, with its
nearest singularity at $t=m_0^2$, $m_0$ the mass of the lightest
tensor.  Because of the mass gap, confinement leads to an exponential
damping, i.e., in the supergravity limit, it again leads to a real
eikonal, but with an exponential cutoff
\be
\chi_{hw} (\tau ,b,z,z') \sim e^{\tau} e^{-m_0b}
\ee
To illustrate the effect of confinement, we show in Fig,
\ref{fig:hw2conf} the ratio of the hard-wall eikonal to the conformal
eikonal calculated numerically, as a function of $b/z_0$ and $z/z_0$,
with $z'\simeq z_0$ fixed. Note the rapid drop when $b>z_0$ and, in
comparison, a relative slow variation in $z$.

\begin{figure}[bthp]
\begin{center}
\includegraphics[width=0.4\textwidth]{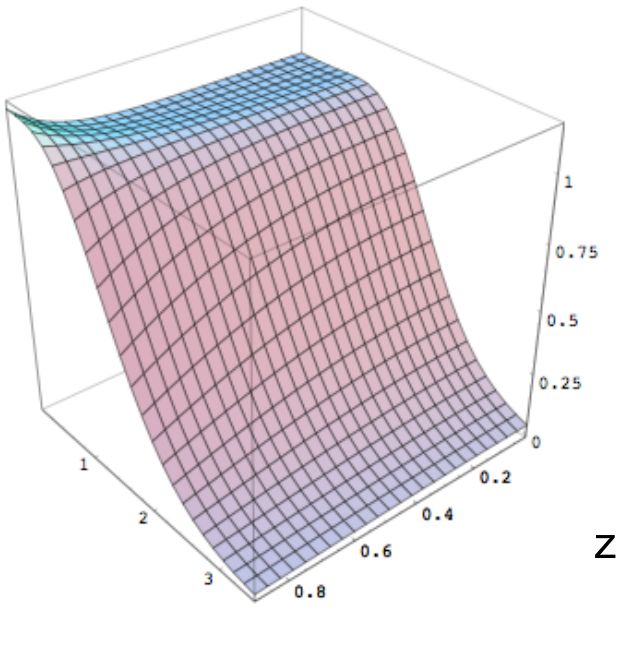}
\caption{Ratio of hard-wall to conformal eikonal Contour in the super-gravity limit,  $\chi_{hw}(\hat s,b,z,z')/\chi_{c}(\hat s,b,z,z')  $. The ratio is independent of $\hat s$ and is plotted against $z/z_0$ and $b/z_0$, with $0<z/z_0< 1$ and $z' = z_0 $ fixed.  }
\label{fig:hw2conf}
\end{center}
\end{figure}

To fully explore the consequence of confinement at finite 't Hooft coupling, it is useful to work with the Pomeron kernel in the $J$-plane. It can be shown that the eikonal for the hard-wall model has a cutoff  at large $b$  of the form
\be
Im\; \chi_{hw}(\tau ,b,z,z')\sim \exp[-  m_1 b  - (m_0-  m_1)^2\;  b^2 / 4 \rho \tau ]  \label{eq:exponential}
\ee
where $m_1$ and $m_0$ are  solutions of       $\partial_z(z^2J_0(mz))\; |_{z=z_0} =0$ and  $\partial_z(z^2J_2(mz))\; |_{z=z_0} =0$ respectively. (We have $m_1\simeq 1.6 \; z_0^{-1}$ and $m_0\simeq 3.8 \;z_0^{-1}$. \cite{Brower:2008cy}). Although the $J$-plane is difficult to carry  out  analytically, it can  be treated, for instance,   numerically.   For our current analysis, we shall   modify our earlier  the conformal hard-wall eikonal, by taking the large-b cutoff, (\ref{eq:exponential}),  into account. For $b$-small, as well as to take into account the proper boundary condition near the IR hard-wall, we shall take $Im \; \chi_{hw}(\tau,b,z,z')$ to be of the form   $Im \;\chi^{(0)}_{hw}(\tau,b,z,z')\sim  Im \;\chi_{c}(\tau ,b,z,z') + {\cal F}(\tau ,z,z')Im \; \chi_{c}(\tau,b,z,z_0^2/z')$.
We therefore adopt the following simple ansatz\footnote{This ansatz should be further improved as one moves to smaller $x$-value beyond the HERA range. This will be addressed in a future publication where we also treat elastic and total cross sections, as well as diffractive Higgs production at LHC.}  where $Im \; \chi_{hw}(\tau,b,z,z')\sim D(\tau ,b)  Im\;  \chi^{(0)}_{hw}(\tau,b,z,z')$, where
\be
D(\tau ,b) =   \left\lbrace
\begin{array}{ll}
1\;, & b< z_0\\
\frac{ \exp[-  m_1 b  -  (m_0-m_1)^2\;  b^2 / 4 \rho \tau ]}{\exp[-  m_1 z_0  - ( m_0-m_1)^2\;  z_0^2 / 4 \rho \tau ]}    \;, & b > z_0 \\
\end{array}
\right .
\label{eq:damping}
\ee
 Lastly, we provide an overall  factor $C(\tau,z,z')$ which can be  fixed by the normalization condition (\ref{eq:ImK hard-wall}).

\begin{figure}[bthp]
\begin{center}

\includegraphics[height=0.65 \textwidth,width=.99\textwidth]{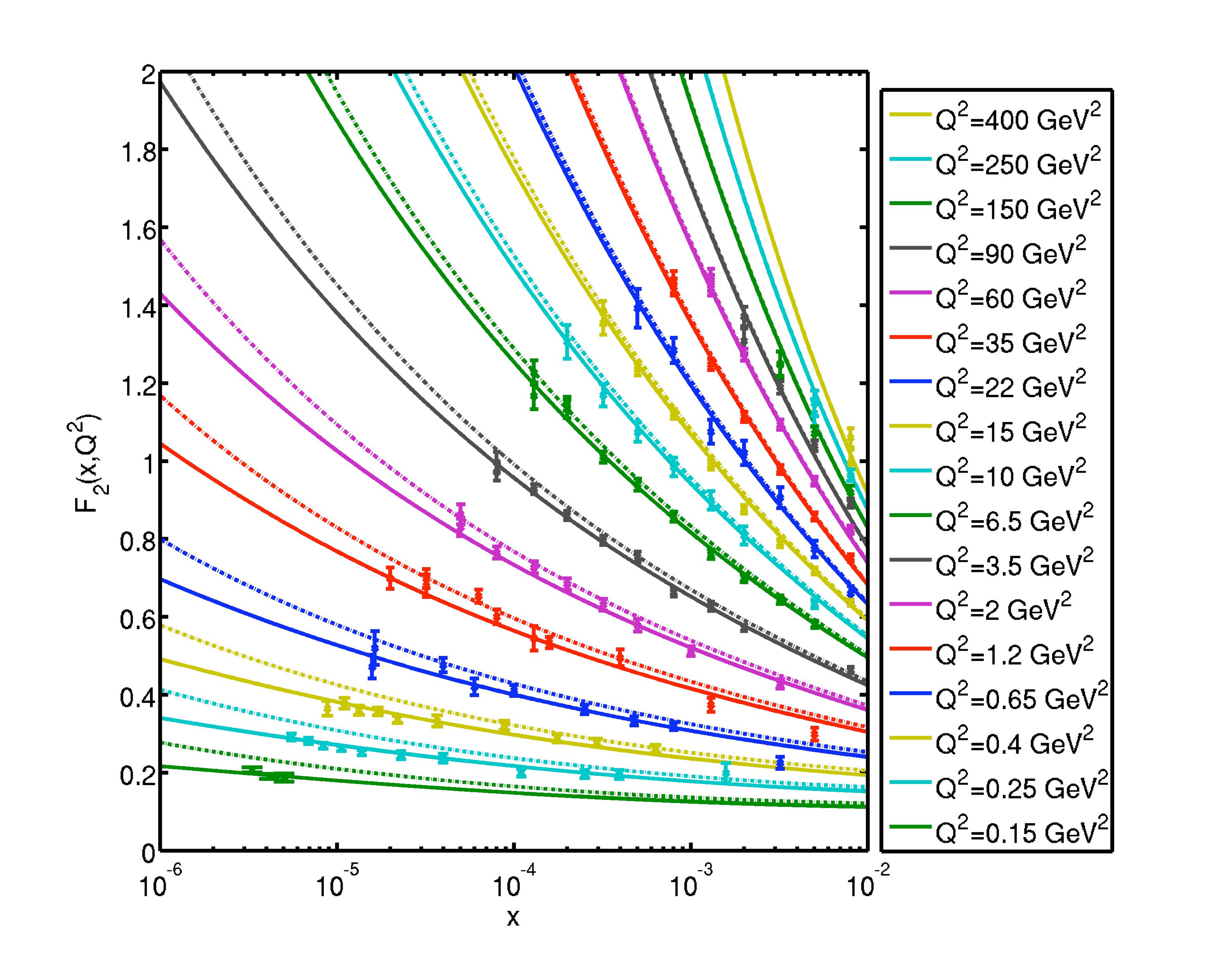}

\caption{Fit to the combined H1-ZEUS small-$x$ data for $F_2(x,Q^2)$ by a hard-wall eikonal treatment. We have exhibited both the hard-wall single Pomeron fits, (in dashed lines), and the hard-wall eikonal, (in solid lines), together for a better visual comparison. The fit include 249 data points, with $x<10^{-2}$, and 34 $Q^2$ values, ranging from $0.1\; GeV^2$ to $400\; GeV^2$.  Only data set for 17 $Q^2$ values are shown.}
\label{fig:combinedhwPomeron}
\end{center}
\end{figure}

In Fig. \ref{fig:combinedhwPomeron}, we have collected both the BPST hard-wall Pomeron fits, (dashed lines), and the hard-wall eikonal, (solid lines), in one place for a better visual effect. For the latter fit, real part of the eikonal is included dispersively. For clarity, we only show data for 17 $Q^2$ sets, out of 34. We find that the confinement-improved eikonal treatment allows a surprisingly good fit to all HERA small-$x$ data, with  $Q^2$ ranging from  $0.1\, GeV^{2}$ to $400\, GeV^{2}$ and for $x<10^{-2}$, with a $\chi^2 = 1.04$, after performing a similar sieve-procedure as done earlier. The parameters for the best fits remain basically the same, with
\begin{equation}
\rho=0.7833\pm 0.0035, \;   g_0^2 =104.81 \pm 1.41 , \; z_{0}=6.04\pm 0.15 \, GeV^{-1}, \;  Q' =0.4439\pm 0.0177 \; GeV.\,   \,\label{eq:parametersB}
\end{equation}
Whereas these two fits are indistinguishable for most HERA data, they do begin to diverge as one moves  to smaller $x$ values, e.g., to the LHC range.

 Our analysis  confirms that saturation effects is minimal for $Q^2\geq O(1) \; GeV^2$ at HERA energy range.  For $Q\leq  O(1) \; GeV^2$,  eikonal treatment can achieve a better fit than that by a single hard-wall Pomeron where saturation effects can begin to be felt.  In Fig.    \ref{fig:Intercepta}, we show by the dotted blue curve  the eikonal improved effective  Pomeron intercept based on hard-wall eikonal. Note that this  further lowers the effective intercept as one moves to smaller $Q^2$.  As mentioned earlier, this is consistent with the view
of ``soft Pomeron'' dominance for hadronic cross sections in the near
forward region, with an effective intercept $\alpha_P\simeq 1.08$.  In Fig. \ref{fig:eikonal}b, we provide a contour plots for this confinement-improved eikonal, $Im\; \chi_{hw}(\hat s, b,z,z')$ at  $b=0$.  By focussing on the eikonal at $b= 0$, we also see that it is increasing important to include non-linear effects, particularly for $Q^2=O(1) GeV^2$. This of course indicates  the onset of saturation, (\ref{eq:saturation}), and this  is  schematically represented in Fig. \ref{fig:HERA_LHC} by a saturation line. In fact, for any $Q^2$, for $1/x$ sufficiently small, eikonalization will always be necessary.

\section{Summary}\label{sec:discussion}

The AdS/CFT correspondence provides  a new approach to deep
inelastic scattering.    In this paper we have carried out  an analysis  of the DIS structure functions at small-x  using the AdS/CFT   correspondence. Our present analysis is based on the work of Brower, Polchinski, Strassler and Tan (BPST)~\cite{Brower:2006ea} where
 the concept of a non-perturbative Pomeron was shown to follow unambiguously for all gauge theories allowing String/Gauge duality.  By identifying deep inelastic scattering
with virtual photon total cross section, this allows a self consistent description
at small-x where the dominant contribution
is the vacuum exchange process.  We find that the BPST Pomeron kernel, along with a  very simple local approximation to
the proton and current wave functions, gives a remarkably good fit
not only at large $Q^2$, dominated by conformal symmetry, but also at small $Q^2$, with an IR hard-wall cut-off of the AdS.

We first
treated  DIS in the small-x limit  to first order in  the conformal approximation limit.  We explain how at  strong coupling   the small-x Regge limit and the large-$Q^2$ limit are unified by discussing the $\Delta-J$ curve and show how the vanishing of anomalous dimension $\gamma_2$ is satisfied automatically. 
We next discuss  the modification due to  confinement, using the hard-wall model as an illustration. We end with a more precise treatment of the effect of confinement on the structure of the eikonal, $\chi(\hat s,b,z,z')$, in the impact space.

This formalism is used to fit the recently combined H1-ZEUS small-x data from HERA~\cite{:2009wt}. We focus   on a single-Pomeron contribution based on a ``local approximation'' for both the current and the proton ``wave functions''.  We first find  that,  at larger $Q^{2}$, e.g., $Q^2\geq  O(1) \; GeV^2$, both the conformally invariant theory and the confined hard-wall model fit the experimental data well, e.g.,  by first restricting to a smaller set of  ZEUS data~\cite{Breitweg:1998dz,Chekanov:2001qu}, for values of $Q^{2}$ ranging from $0.65\, GeV^{2}$ to $650\, GeV^{2}$ and for $x<10^{-2}$. Armed with this initial success, we  next apply our results  to  the  combined H1-ZEUS small-x data. This is a much larger data set, and, in particular, the set now extends to much smaller $Q^2$ values. We  find that the confinement-improved treatment (hard-wall model) allows a surprisingly good fit to all HERA small-x data, with  $Q^2$ ranging from  $0.1\, GeV^{2}$ to $400\, GeV^{2}$ and for $x<10^{-2}$, with a $\chi_{d.o.f.}^2 = 1.07$, and best fits to various parameters given by (\ref{eq:parameters}), e.g., with a BPST intercept at $j_0\simeq 1.22$.  In particular, we find that the $Q^2$-dependence for $\epsilon_{eff}$ observed at HERA, Fig. \ref{fig:Intercepta}, can be attributed primarily to diffusion  for $Q^2$ large and to confinement effects for $Q^2$ small.  In contrast, the conformal fits fails when the low-$Q^2$ data is included. The single-Pomeron hard-wall fit  also  indicates possible onset of ``saturation'' for small $Q^2$, e.g., for $Q^2 \leq O(1) \; GeV^2$.

Finally we carried out a nonlinear eikonal analysis.   It  is  now important to fully explore the dependence of the eikonal, $\chi( s, \vec b, z,z')$, on the 3-dimensional transverse space, i.e., $\vec b$ and $z$.  For the conformal limit, this  is given by Eq. (\ref{eq:ImK conformal}). For the hard-wall model, a more elaborate treatment is required.  Due to a much stronger exponential cutoff in the impact parameter, confinement modifies drastically the conformal result.  The scale of the cutoff is set by the lowest tensor glueball mass, which in turn depends on the confinement scale. Our analysis  confirms that saturation effects are small for $Q^2\geq O(1) \; GeV^2$ at HERA energy range. However, for $Q\leq  O(1) \; GeV^2$, the conformal-eikonal treatment remains inadequate. In contrast, confinement-improved eikonal treatment allows an improved  fit to all HERA small-x data,  with a $\chi_{d.o.f.}^2 = 1.04$ and best fits to various parameters given by (\ref{eq:parametersB}).

Surprisingly, we find that  confinement effects persist at an increasingly large value of  $Q^2$  as $1/x$ increases, as indicated schematically by a confinement line  in Fig. \ref{fig:HERA_LHC}. A confinement-improved BPST Pomeron treatment  allows a surprisingly good fit to all HERA small-x data.  Nonlinear effect due to eikonalization  is small but begins to be noticeable for low-$Q^2$  HERA data at small-x, indicating imminent  approach of saturation. This is also represented schematically in Fig. \ref{fig:HERA_LHC} by a saturation line. Note that saturation line lies above the confinement line, indicating that the physics of saturation should be discussed in a confining setting.  Clearly,  this observation is of significance for diffractive central production of jets, Higgs, et al. at LHC.  Equally important is the application to the study  on the onset of Froissart-like behavior~\cite{Block:2006dz,Berger:2007vf,Block:2005pt}  for DIS as well as the possible extension to a study on the ultra-high energy neutrino scattering and for the experimental search for extra-galactic neutrinos \cite{Gandhi:1998ri,Berger:2007ic}. Applications to these processes will be reported separately.  Lastly, some other recent discussions related to  DIS from AdS/CFT can also be found in \cite{Pire:2008zf,Yoshida:2009dw,Gao:2009ze,Hatta:2009ra,Bartels:2009sc,Bartels:2009jb,Gao:2009se,Marquet:2010sf,Iancu:2010wy}.

\vskip40pt

\noindent {\underline{Acknowledgments:}} 
We are pleased to acknowledge useful conversations
with M. Block, E.  Levin and C. Vergu.  The
work of R.B. was supported by the Department of Energy under
contract~DE-FG02-91ER40676,
that of I.S.  by the Department of Energy
under contracts DE-FG02-04ER41319 and DE-FG02-04ER41298, and that of M.D. and  C.-IT.  by the
Department of Energy under contract~DE-FG02-91ER40688, Task-A, 
I.S. would like to thank the Brown High Energy Theory Group for visit  where this work was initiated. R.B. and C.-I.T. would like to thank the Aspen Center
for Physics for its hospitality during the completion of this work.

\newpage
\bibliographystyle{utphys}
\bibliography{DISAdS}
\end{document}